\documentclass{ws-procs9x6}

\begin{document}
\markboth{David L Wiltshire}
{Dark energy without dark energy}
\font\fiverm=cmr5\font\sevenrm=cmr7\font\sevenit=cmmi7
\def\ns#1{_{\hbox{\sevenrm #1}}}
\def\dd{{\rm d}}\def\OM{\mean\Omega}\def\ab{\mean a}\def\rhb{\mean\rh}
\def\ds{\dd s} \def\e{{\rm e}} \def\etal{{\em et al}.}
\def\ga{\gamma}\def\de{\delta}\def\bq{\mean q}\def\bT{\mean T}
\def\th{\theta}\def\Th{\vartheta}\def\ph{\phi}\def\rh{\rho}\def\si{\sigma}

\def\mean#1{{\vphantom{\tilde#1}\bar#1}}
\def\bH{\mean H}\def\gb{\mean\ga}\def\gc{\gb\Z0}
\def\goesas{\mathop{\sim}\limits}

\def\w#1{\,\hbox{#1}} \def\Deriv#1#2#3{{#1#3\over#1#2}}
\def\Der#1#2{{#1\hphantom{#2}\over#1#2}} \def\br{\hfill\break}
\def\ts{t} \def\tc{\tau} \def\tv{\tc\ns{v}} \def\tw{\tc\ns{w}}
\def\Dts{\mathop{\hbox{$\Der\dd\ts$}}}
\def\Dtc{\mathop{\hbox{$\Der\dd\tc_{\!\lower1pt\hbox{\fiverm w}}$}}}
\def\rhcr{\rh\ns{cr}}
\def\Z#1{_{\lower2pt\hbox{$\scriptstyle#1$}}}
\def\X#1{_{\lower2pt\hbox{$\scriptscriptstyle#1$}}}
\def\Ns#1{\Z{\hbox{\sevenrm #1}}}
\def\av{{a\ns{v}\hskip-2pt}} \def\aw{{a\ns{w}\hskip-2.4pt}}
\def\etw{\eta\ns w} \def\etv{\eta\ns v}
\def\FF{{\cal F}} \def\Fi{\hbox{\footnotesize\it fi}}
\def\Hh{H} \def\Hm{H\Z0} \def\frn#1#2{{\textstyle{#1\over#2}}}
\def\half{\frn12} \def\DD{{\cal D}} \def\Vav{{\cal V}}
\def\lsim{\mathop{\hbox{${\lower3.8pt\hbox{$<$}}\atop{\raise0.2pt\hbox{$\sim$}}
$}}}
\def\dL{d\Z L} \def\dA{d\Z A} \def\DE{\Delta}
\def\kmsMpc{\w{km}\;\w{sec}^{-1}\w{Mpc}^{-1}}
\def\ab{{\bar a}} \def\LCDM{$\Lambda$CDM}
\def\OmMn{\Omega\Z{M0}}\def\Hb{\bH\Z{\!0}}\def\etBg{\eta\Z{B\ga}}
\def\OMBn{\mean\Omega\Z{B0}}\def\OmBn{\Omega\Z{B0}}
\def\finfty{\mathop{\hbox{\it fi}}} \def\etb{\mean\eta} \def\dOM{\dd\Omega^2}
\def\bx{{\mathbf x}}\def\fvi{{f\ns{vi}}} \def\fwi{{f\ns{wi}}}
\def\ave#1{\langle{#1}\rangle}\def\Rav{\ave{\cal R}}\def\gd{{{}^3\!g}}
\def\QQ{{\cal Q}}\def\rhav{\ave\rh}\def\pt{\partial}
\def\FF{{\cal F}} \def\FI{\FF\Z I} \def\kv{k\ns v} \def\fv{{f\ns v}}
\def\OMM{\OM\Z M}\def\OMk{\OM\Z k}\def\OMQ{\OM\Z{\QQ}}
\def\fvf{\left(1-\fv\right)}\def\rw{r\ns w} \def\OmM{\Omega\Z M}
\def\epi{\epsilon_i} \def\OMMn{\OM\Z{M0}}\def\fvn{f\ns{v0}}
\def\beq{\begin{equation}} \def\eeq{\end{equation}}
\def\bea{\begin{eqnarray}} \def\eea{\end{eqnarray}}
\def\ApJ#1{Astrophys.\ J.\ {\bf#1}} \def\PL#1{Phys.\ Lett.\ {\bf#1}}
\def\AsJ#1{Astron.\ J.\ {\bf#1}}
\def\CQG#1{Class.\ Quantum Grav.\ {\bf#1}}
\def\GRG#1{Gen.\ Relativ.\ Grav.\ {\bf#1}}
\def\AaA#1{Astron.\ Astrophys.\ {\bf#1}}
\def\PRL#1{Phys.\ Rev.\ Lett.\ {\bf#1}} \def\PR#1{Phys.\ Rev.\ {\bf#1}}
\def\MNRAS#1{Mon.\ Not.\ R.\ Astr.\ Soc.\ {\bf#1}}
\def\ries{\cite{Riess06}}
\title{DARK ENERGY WITHOUT
DARK ENERGY\footnote{Based
on talks presented at the NZIP2007, GRG18 and Dark2007 conferences.}}
\author{David L. Wiltshire$^\dag$}
\address{Department of Physics and Astronomy, University of Canterbury,\\
Private Bag 4800, Christchurch 8140, New Zealand\\
$^\dag$E-mail: David.Wiltshire@canterbury.ac.nz\\
http://www2.phys.canterbury.ac.nz/$\goesas$dlw24/ }


\begin{abstract}
An overview is presented of a recently proposed ``radically conservative''
solution to the problem of dark energy in cosmology. The proposal
yields a model universe which appears to be quantitatively viable, in
terms of its fit to supernovae luminosity distances, the angular scale of the
sound horizon in the cosmic microwave background (CMB) anisotropy spectrum,
and the baryon acoustic oscillation scale. It may simultaneously resolve key
anomalies relating to primordial lithium abundances, CMB ellipticity, the
expansion age of the universe and the Hubble bubble feature. The model uses
only general relativity, and matter obeying the strong energy condition, but
revisits operational issues in interpreting average
measurements in our presently inhomogeneous universe, from first principles.
The present overview examines both the foundational issues concerning the
definition of gravitational energy in a dynamically expanding space, the
quantitative predictions of the new model and its best--fit cosmological
parameters, and the prospects for an era of new observational tests in
cosmology.
\end{abstract}
\bodymatter
\section{Introduction}

Dark energy is widely described as the biggest problem in cosmology
today; one which may demand new physics and a theoretical paradigm
shift\cite{dark1}.
In this paper I suggest that the solution to the problem of dark
energy is intimately related to the correct understanding of
observational anomalies -- in particular, the observed abundance and
emptiness of voids, which has also elicited separate calls for a
paradigm shift\cite{P_void}. I propose that the paradigm shift that is
required to understand both these issues actually entails no ``new''
physics, but a revisiting of fundamental issues relating to the subtlety of
the definition of gravitational energy within general relativity, and its
relation to the careful modelling of the distribution of matter that we
actually observe.

The punchline is that cosmic acceleration can be understood as an apparent
effect, and dark energy as a misidentification of those aspects of
cosmological gravitational energy which by virtue of the strong equivalence
principle cannot be localized\cite{opus}, namely gradients in the quasilocal
gravitational energy associated with spatial curvature gradients, and
the kinetic energy of expansion, between bound systems
and the volume--average position in freely expanding space. With this
interpretation, a two--scale model can be constructed\cite{opus},
and a simple exact solution\cite{sol} of the relevant equations of cosmic
evolution, the Buchert equations\cite{buch1}, can be found. This
solution provides a fit to type Ia supernovae (SneIa) luminosity
distances which is statistically indistinguishable from that
of the standard Lambda Cold Dark Matter (\LCDM) model, while
simultaneously satisfying key independent cosmological tests
and offering the potential to resolve significant observational
anomalies\cite{LNW}.

Over the past decade we have come to think of ``dark energy'' as
a homogeneous isotropic fluid--like quantity, with a local pressure
related to its energy density by $P=w\rho$, where $w<{-1}/3$, so that
the strong energy condition is violated. In fact, observations appear
to be most consistent with a pure vacuum energy, $w=-1$, and
determination of the value of the parameter, $w(z)$, as a function
of redshift, $z$, is seen as the goal of ``precision cosmology''.
What is proposed here, if correct, will turn this situation on its
head. We can look forward to an era of precision cosmology, but one in
which the focus will be on the complex hierarchical structure
of the universe, rather than any one simple fluid equation of state.
Nonetheless, while the proposed solution is intimately connected to the
growth of inhomogeneities, and their backreaction on the
geometry of the universe \cite{brpapers}, at its heart it
addresses the question of the normalization of gravitational
energy relative to observers within the observed structure. Thus I claim
that the solution to the central foundational question
does concern energy, and since ``nothing'' is ``dark'' the terminology
``dark energy'' is actually quite apt for the new proposal, if the community
will allow the liberty of a change to the assumed definition of those words.

In this paper I will give an overview of the proposed solution
to the problem of dark energy, the extent of its present quantitative
successes, and more importantly the directions for future work.
I will present fewer technical details
than may be found in papers already published\cite{opus,sol,LNW}
or in preparation\cite{equiv,obs}. I aim to give a general overview to
researchers in astrophysics, particle physics or general relativity,
who have no specific prior experience with the averaging of inhomogeneous
cosmological models.

\section{The universe we observe}

Our most widely tested ``concordance model'' of the universe relies on the
assumption of an isotropic homogeneous geometry at all epochs of cosmic
evolution. By the evidence of the cosmic microwave background (CMB) radiation,
the universe was very smooth at the time of last scattering, and these
assumptions were completely well--justified then. The departure from
homogeneity was of order $\de\rh/\rh\goesas10^{-5}$ in photons and the baryons
that couple to them, and perhaps of order $\de\rh/\rh\goesas10^{-3}$ in
non--baryonic dark matter, which gives rise to the potential wells
responsible for the dominant Sachs--Wolfe effect. By the Copernican principle,
the assumption of global isotropy and homogeneity is completely justified
at the epoch of last scattering, and it is safe to assume that the
evolution of the universe was therefore extremely closely modelled
by the Friedmann--Lema\^{\i}tre--Robertson--Walker (FLRW) solutions
at that epoch. Furthermore, if we consider the spectrum of CMB
anisotropies \cite{wmap1,wmap3} then overall it appears that globally
the universe is very close to spatially flat. Its initial evolution
at the time of last scattering would therefore have been very close
to that of an Einstein--de Sitter model at that epoch, even in the
case of the \LCDM\ paradigm since ``dark energy'' only becomes
appreciable at late epochs.
\begin{figure}
\psfig{file=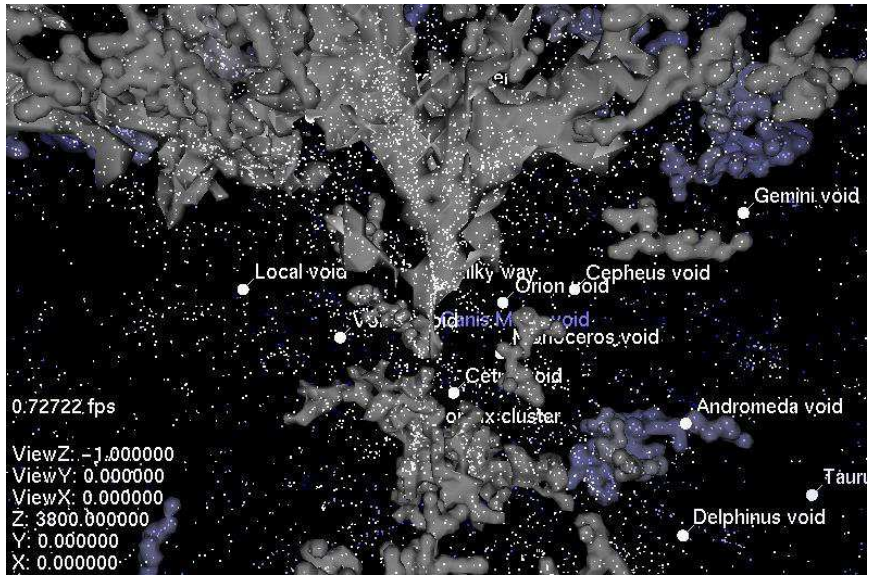,width=4.5in}
\caption{Local voids and bubbles from the 6df survey. Courtesy of
A.~Fairall.}
\label{fig1}
\end{figure}

At the present epoch, however, the distribution of matter is far from
homogeneous on scales less than 150--300 Mpc. The actual universe has a
sponge--like structure, dominated by huge voids. These voids are surrounded
by bubble walls, and threaded by filaments, within which clusters of galaxies
are located. Recent surveys suggest\cite{HV} that some 40--50\% of the present
volume of the universe is in voids of a characteristic scale 30$h^{-1}$ Mpc,
where $h$ is the dimensionless Hubble parameter, $\Hm=100h\kmsMpc$.
If smaller minivoids \cite{minivoids} and larger supervoids \cite{bigvoid}
are included, then our observed universe is presently void--dominated
by volume.

Quite apart from the fact that this observed structure appears emptier
than the vistas that Newtonian $N$--body simulations typically produce,
the mere fact that the universe is presently inhomogeneous means that the
assumptions implicit in the FLRW approximation can no longer be justified
at the present epoch in the almost exact sense that they were justified at
the epoch of last scattering. Homogeneity only applies at the present day
in an {\em average sense}. The manner in which we take the average, and the
operational issues associated with this are not trivial, since the problem
of fitting the local geometry of bound systems to the full dynamical
geometry of the evolving universe \cite{fit1,fit2} is a complicated
one.

Since a broadly isotropic Hubble flow is observed, an almost FLRW geometry
would appear to be a good approximation at some level of averaging,
if our position is a typical one. I will demonstrate, however, that
the assumption that the observed universe evolves exactly as a
smooth featureless dust fluid means that we factor out central physically
important questions which need to be understood to correctly relate
our own observations to the average geometry.

\subsection{The Sandage--de Vaucouleurs paradox}
There is a central foundational paradox concerning the expansion of
the universe, which others have called the ``Hubble--de Vaucouleurs
paradox'' \cite{paradox1,bary} or the ``Hubble--Sandage paradox''
\cite{paradox2}, but which I will call the {\em Sandage--de Vaucouleurs
paradox} since it was originally raised by Sandage and collaborators
\cite{STH} in objection to de Vaucouleurs' hierarchical cosmology
\cite{deVau}.

The problem is that in the conventional way of thinking about
cosmological averages, below the scale of homogeneity we should expect large
statistical scatter in the peculiar velocities of galaxies. In fact, if we
were to average on scales of order 20Mpc, which is about 10\% of the scale
of homogeneity, then the statistical scatter should be so large that
no linear Hubble law can be extracted. Yet 20Mpc is the scale on which
Hubble originally extracted his famous linear law. By conventional
understanding the statistical quietness of the local Hubble flow
does not make sense.

One might attempt to explain the Sandage--de Vaucouleurs paradox as
a consequence of dark energy, since it is well--known that in any
FLRW model which expands forever -- with or without dark energy -- peculiar
velocities eventually decay. However, if one tries to numerically model
12.5 Gyr of motion of very local galaxies, then it turns out that the
initial conditions of each galaxy appear to have the most bearing on
the problem\cite{paradox2}. In particular, one can fit a realistic motion
with a cosmological constant, $\Lambda$, or alternatively in an open
universe without $\Lambda$. Similar conclusions are also reached in
constrained dark matter simulations of the Local Group \cite{hoffman}.
Furthermore, the \LCDM\ parameters
required for the velocity dispersion predicted by structure formation
to match the observed velocity dispersion, do not coincide with the
concordance parameters\cite{AP}. Evidently a cosmological constant
alone cannot explain the quietness and linearity of the local
Hubble flow.

A related, but not exactly equivalent, issue is that using a conventional
kinematic approach, the peculiar velocities of local galaxies appear to be
at least a few factors too small to have arisen from a smooth distribution
of matter at the time of recombination\cite{wh1}.

\section{Averaging, backreaction and dark energy: the debate}

The present distribution of matter is clearly very complex, and since
we cannot solve the Einstein equations for this distribution of matter
analytically, there is an important question as to how we operationally
match the average geometry of this distribution to the simple FLRW models
that we know how to solve.
Given that the nonlinear growth of structure appears to be roughly correlated
to the epoch when cosmic acceleration is inferred to begin,
a number of cosmologists have questioned whether the
introduction of a smooth dark energy is a mistaken interpretation
of the observations \cite{brpapers}. Attention has focused on the
possibility that effects attributed to cosmic acceleration may
actually be due to the backreaction from
the growth of inhomogeneities in determining the geometry of the
observed universe, without exotic dark energy. Different
interpretations of a host of complex technical issues have led to a vigorous
debate.

There are two large streams of research, which I will not focus on, for
physical reasons. The first is backreaction in perturbation
theory about an FLRW background, which became the focus of debate
in 2005 following public attention generated by papers of Kolb
and collaborators\cite{Kolb1,Kolb2}. Within perturbation theory one may
demonstrate that there is a potentially significant effect from
backreaction\cite{Kolb2}. However, this argument cannot be conclusive -- if
the second--order terms in the perturbation expansion exhibit an effect which
might be interpreted as cosmic acceleration, such an effect may go away when
the third--order terms are considered, and so on. Perturbation theory
is very relevant near the epoch of last scattering, when the assumption
of homogeneity was extremely good; but by the present epoch the
nonlinear structures are so numerous and complex that we are beyond the
regime of its applicability. Thus perturbation
theory cannot provide a complete solution, and will not be discussed here.

The second stream of research I will not consider are those that involve
exact inhomogeneous solutions of the Einstein equations, at the expense
of introducing matter distributions which are unlikely. The spherically
symmetric Lema\^{\i}tre--Tolman--Bondi (LTB) solutions are perhaps the
most well--studied class of such models. While they may be very realistic
descriptions of single voids,
to apply them to the universe as a whole violates the Copernican principle,
which I shall retain. One may obtain LTB solutions within which
one can obtain reasonable fits to supernovae luminosity
distances\cite{LTBmatch}. However, in my view, given their high degree
of symmetry, these are at best toy models,
which one cannot hope to reproduce in structure formation scenarios
based on our knowledge of the power spectrum of density perturbations
at the time of last--scattering.

To confront the actual inhomogeneous universe, which has no particular
spatial symmetries below the scale of homogeneity, we must deal with schemes
that average the full non--linear Einstein equations. There are many
schemes for constructing averages, including those of
Zalaletdinov\cite{Zal} and Buchert\cite{buch1}. There are various
grounds for favouring the approach of Zalaletdinov\cite{Zal}, which is
fully covariant and averages all of the Einstein equations. However,
Zalaletdinov's scheme is a general one, and for the cosmological
problem additional assumptions are required. In Buchert's scheme one
just average scalar quantities, and an additional integrability condition
is required for the equations to close. However, with reasonable
cosmological assumptions, the correlation tensor in Zalaletdinov's scheme
takes the form of a spatial curvature \cite{CPZ}, and Buchert's scheme
can be realized as a consistent limit \cite{PS}. Furthermore
Buchert's scheme yields equations which are close to the Friedmann
equations. Given that the Friedmann equations have worked so well to
date, Buchert's average would appear to give a natural framework within
which corrections to the Friedmann evolution can be quantitatively
examined for the universe we observe. I shall adopt Buchert's scheme,
with caveats to be discussed.

Buchert's scheme strictly deals with irrotational dust cosmologies,
characterized by an energy density, $\rh(t,\bx)$, expansion, $\Th(t,\bx)$,
and shear, $\si(t,\bx)$, on a compact domain, $\DD$, of a suitably defined
spatial hypersurface of constant average time, $t$, and spatial 3--metric,
$^3g_{ij}$. Angle brackets are taken
to denote the spatial volume averages, e.g., for the scalar curvature
$$\Rav\equiv\left(\int_\DD\dd^3x\sqrt{\det\gd}{\cal R}(t,\bx)\right)/\Vav(t)\,
$$
with $\Vav(t)\equiv\int_\DD\dd^3x\sqrt{\det\gd}$.
The important lesson of Buchert averaging is that time evolution and
averaging to do not commute\cite{buch1}. Generally for any scalar $\Psi$,
\beq
\Deriv{\dd}t{}\ave\Psi-\ave{\Deriv{\dd}t\Psi}=\ave{\Psi\Th}-\ave\Th\ave\Psi
\label{noncom}\eeq
The fact that the r.h.s.\ of (\ref{noncom}) does not vanish, as is the
case for the FLRW cosmologies, is a manifestation of {\em backreaction}.

Applied to the equations of cosmic evolution one obtains the exact
{\em Buchert equations}
\bea
3{\dot\ab^2\over\ab^2}&=&8\pi G\ave\rh-\half\Rav-\half\QQ,\label{buche1}\\
3{\ddot\ab\over\ab}&=&-4\pi G\ave\rh+\QQ,\label{buche2}\\
\pt_t\ave\rh&+&3{\dot\ab\over\ab}\ave\rh=0,
\label{buche3}\eea
where $\ab(t)\equiv\left[\Vav(t)/\Vav(t\Z0)\right]^{1/3}$,
\beq
\QQ\equiv\frac23\left(\langle\Th^2\rangle-\langle\Th\rangle^2\right)-
2\langle\si\rangle^2
\label{backr}\eeq
and following integrability condition follows from
(\ref{buche1})--(\ref{backr}):
\beq \pt_t\left(\ab^6\QQ\right)+\ab^4\pt_t\left(\ab^2\Rav\right)=0.\eeq
It is observed that the kinematic backreaction term, $\QQ$, enters
(\ref{buche2}) with the same sign as a cosmological constant in the
equivalent Raychaudhuri equation in the FLRW paradigm, but enters
(\ref{buche1}) with the {\em opposite} sign to a cosmological constant in
the Friedmann equation. Given this fact, together with the domain-dependence
of the average, the question of whether Buchert's scheme can generate
sufficient backreaction to give apparent acceleration for realistic initial
conditions has been the subject of much debate. On reasonable grounds one
might conclude\cite{IW} that while the effects are real, they are too small in
magnitude to give departures sufficiently large from the FLRW expectation to
register as cosmic acceleration.

It is at this point in the argument caution must be exercised. As
every student of general relativity knows, one cannot simply write down
a time parameter and assume that it is the parameter one measures on
one's clock, without specifying how it is related to local invariants.
We actually measure luminosity {\em distances}, and the deduction of cosmic
acceleration involves two time derivatives. We therefore
have to be extremely careful to operationally specify how $t$ is to be
related to our own clocks. It must be observed that the scale--factor
$\ab(t)$ is not related to an exact local metric, that has been substituted
in Einstein's equations. Rather we must solve the Buchert equations, and
determine how the best--fit almost--FLRW geometry that is obtained
relates operationally to our own measurements.

\section{Finite infinity and gravitational energy}

In a completely arbitrary inhomogeneous universe the calibration of rods and
clocks at one point relative to another can vary arbitrarily. However,
our clocks and rods and those of stars in distant galaxies we observe,
appear to a very good approximation to be determined by geodesics of ideal
solutions -- the Kerr and Schwarzschild geometries -- in which space is
asymptotically flat, with an exact time symmetry at spatial infinity.
This time symmetry, mathematically described by a timelike Killing
vector, must be an idealization which breaks down at some level, since
the universe is expanding.

In his pioneering work on the fitting problem, Ellis\cite{fit1}
suggested its solution should involve the notion
of {\em finite infinity}\footnote{As finite
infinity adds a new notion of infinity to the concepts timelike,
spatial and null infinity, a new mathematical symbol, $\finfty$, is
appropriate - in \LaTeX: $\rm\backslash def\backslash finfty\{\backslash
mathop\{\backslash hbox\{\backslash it\ fi\}\}\}$.}, ``$\finfty\,$'',
namely a timelike surface within which the dynamics of an isolated
system such as the solar system can be treated without reference to the
rest of the universe. After all, the matter in our galaxy and other typical
galaxies broke away from the Hubble flow to be come a bound system well over
10 billion years ago. With sufficient computing resources, we can
integrate the motions of stars within our galaxy for billions of years
without considering the dynamics of the universe outside the galaxy.
Thus within finite infinity spatial geometry might be considered to be
effectively asymptotically flat, and governed by ``almost'' Killing
vectors.

The concept of finite infinity does not appear to have been further
developed in the intervening two decades. However, given that the
normalization of our clocks in the idealized Schwarzschild and Kerr
geometries is related to the timelike Killing vector at spatial infinity,
finite infinity would appear to be the appropriate reference surface
for the definition of gravitational energy, which in stationary
spacetimes is tied to the asymptotically flat region\cite{ge1}.
In Newtonian terms, it is the scale at which we set the zero of the
gravitational potential.

The definition of gravitational energy in general is an extremely difficult
problem, on account of the fact that space itself is dynamical and
can carry energy and momentum. By the strong equivalence principle, since
the laws of physics must coincide with those of special relativity at a
point, it is only internal energy that can be localized in an
energy--momentum tensor on the r.h.s.\ of the Einstein equations. Any uniquely
relativistic aspects of gravitational energy associated with spatial
curvature and geometrodynamics cannot be included in the energy momentum
tensor, but are at best described by a quasilocal formulation\cite{quasi}.

Einstein himself worried about the problem of quasilocal gravitational
energy, in terms of energy--momentum complexes, and many mathematical
relativists since Einstein have also considered the problem. It is quite
possible that the general problem of a definition of quasilocal energy
does not have a solution, since it depends on the split of space and
time. Here I will not be interested in the general problem for an
arbitrary solution of the Einstein equations, but in the specific problem
for a universe which was effectively homogeneous and isotropic at
last scattering, with nearly scale--invariant perturbations.

The fact that quasilocal energy is not part and parcel of theoretical
framework of the FLRW cosmology, is easily illustrated by the Newtonian
perspective \cite{Bondi1}.
The l.h.s.\ of the standard Friedmann equation with $\Lambda=0$ can be
regarded as the difference of a kinetic energy density per unit rest mass,
$E\ns{kin}=\dot\ab^2/(2\ab^2)$ and a total energy density per unit
rest mass $E\ns{tot}=-k/(2\ab^2)$ of the opposite sign to the
Gaussian curvature, $k$, while the energy--momentum tensor on the r.h.s.\
is the Newtonian potential energy per unit rest mass. The terms in the Einstein
tensor represent forms of gravitational energy, but since they are identical
for all observers in an isotropic homogeneous geometry, one can always
synchronize clocks and calibrate rods of ideal isotropic observers
unambiguously.

As soon as inhomogeneity enters the game, however, one will have
gradients in both the kinetic energy of expansion, and in spatial curvature.
Given the present value of the Hubble constant, it is likely that
the spatial curvature gradient is the most significant effect.
Quasilocal energy gradients between a bound system and finite infinity
could conceivably be relevant for asymptotic galactic dynamics, or
even the solution of the Pioneer anomaly.
This speculation is left for future work. The proposal of Refs.\
[\refcite{opus,sol}] is concerned principally with dynamically evolving
spatial curvature gradients. In this context, one should observe that
the idea that negative spatial curvature is associated with positive
gravitational energy, evident already in the Newtonian framework,
remains true in the LTB models, where the Gaussian curvature is
replaced by an inhomogeneous energy function $E(r)$.

Cosmological gravitational energy is largely uncharted territory in the
more rigorous quasilocal framework \cite{quasi} within which the problem
should ultimately be framed. In any quasilocal formulation results depend
crucially on the reference spacetime and surface of integration. Recently
Chen, Liu and Nester\cite{CLN} have obtained a result over which they expressed
surprise, but which is consistent with the present proposal. They find that
for an isotropic observer in synchronous gauge in a $k=-1$ Friedmann universe
the quasilocal energy in their particular Hamiltonian formalism is negative.
A similar result is obtained using a different approach by Garecki\cite{Ga}.
These results are expected in the current approach, since one is
effectively subtracting a fiducial flat spacetime in each case, and the
{\em relative} sign of energy depends on the observer. An isotropic $k=0$
Friedmann observer has zero quasilocal energy in the approach of Chen, Liu and
Nester; thus relative to the $k=-1$ geometry the $k=0$ geometry has negative
quasilocal energy, but conversely relative to the $k=0$ geometry the $k=-1$
has positive quasilocal energy. Our viewpoint here will be that the fiducial
reference point is the $k=0$ geometry of the finite infinity region. This
agrees with the Newtonian version of energy in the Friedmann equation, the
LTB energy function, and with the idea that binding energy is
negative.\cite{ge1}

\section{Finite infinity and the true critical density}

In the standard FLRW cosmology the critical density is proportional to
the square of the Hubble parameter. However, in an inhomogeneous
cosmology this is no longer true,
\beq\rhcr\ne{3H\ns{av}^2\over8\pi G}\label{Feq}\eeq
as the average Hubble parameter, $H\ns{av}$, is related to the spatial
scale of averaging and the clock chosen. In the spirit of the spherical
collapse model, we are familiar with the idea that a perturbation near
the time of last scattering can evolve as an FLRW model of different
density, until it turns around and collapses, by which point other
regions closer to the true mean density are within the past light cone.

In a universe which grew from a nearly scale--invariant spectrum of
perturbations, there will always be spatial correlations of some density
perturbation which is relatively far from the mean, once one samples
on scales close to the spatial scale of the particle horizon. This
is a simple consequence of cosmic variance, and is illustrated by
the following analogy. Take a bucket filled with grains of sand of
identical diameter, whose density is Gaussian distributed about a mean.
Provided the container is much larger than the individual grains then the
mean density of the bucket can be expected to be extremely close to the mean
density of all the sand from the beach on which it was collected.
If one repeats the exercise for stones of larger and larger diameter,
the mean density of the bucket will on average differ more and more
from the mean density of the beach as the diameter of the stones becomes
comparable to the size of the bucket, by a $\sqrt{N}$ statistic.

Cosmic variance\footnote{I use the term {\em cosmic variance} in a general
sense, rather than restricting it to one statistical aspect of the CMB
temperature fluctuations which arise as an observable consequence of the
variance in the density perturbation spectrum.} in the actual universe is
complicated by three things. Firstly, since general relativity is causal, the
bucket in the analogy is the past light cone which grows with time. Secondly,
the perturbations are not uniform density but contain other perturbations
embedded within them, like Russian dolls. Thirdly, below a scale of
apparent homogeneity the initial perturbations can no longer be thought of
as perturbations. They have undergone a non-linear transition to
form the structure that we see: stars, globular clusters, and clusters of
galaxies strung in filaments and bubbles around voids.

Every inflationary cosmologist who has thought about the foundations
of the subject realises that the present density of the observable universe
need not be the same as the whole universe, observable and
unobservable, and in general these should differ on account of cosmic
variance in a scale--free density perturbation spectrum. Nonetheless, in
the FLRW paradigm we unwittingly end up violating this. As long as we
assume that the average density evolves by the Friedmann equation, we
implicitly assume that the mean density in our past light cone today is
the same as the whole ensemble of observable universes. As long as we
evolve the average geometry purely by the Friedmann equation, with
equality in (\ref{Feq}), we will miss an important aspect of
actual cosmic evolution. Furthermore, as long we study inhomogeneity
in the context of a Swiss cheese model \cite{gruyere} by cutting and
pasting holes into a background which still evolves by the Friedmann
equation alone, we will still fail to uncover essential features of cosmic
evolution.

There is perhaps no obvious way to define finite infinity
in an arbitrary inhomogeneous background. To proceed I make the crucial
observation that since our universe was effectively homogeneous and
isotropic at last scattering, a notion of a universal critical density
scale did exist then. It was the density required for gravity
to overcome the initial uniform expansion velocity of the dust fluid.
I will assume, as is consistent with primordial inflation, that the present
horizon volume of the universe was very close to the critical density
at last scattering, with scale--invariant perturbations.

\begin{figure}
\centerline{\psfig{file=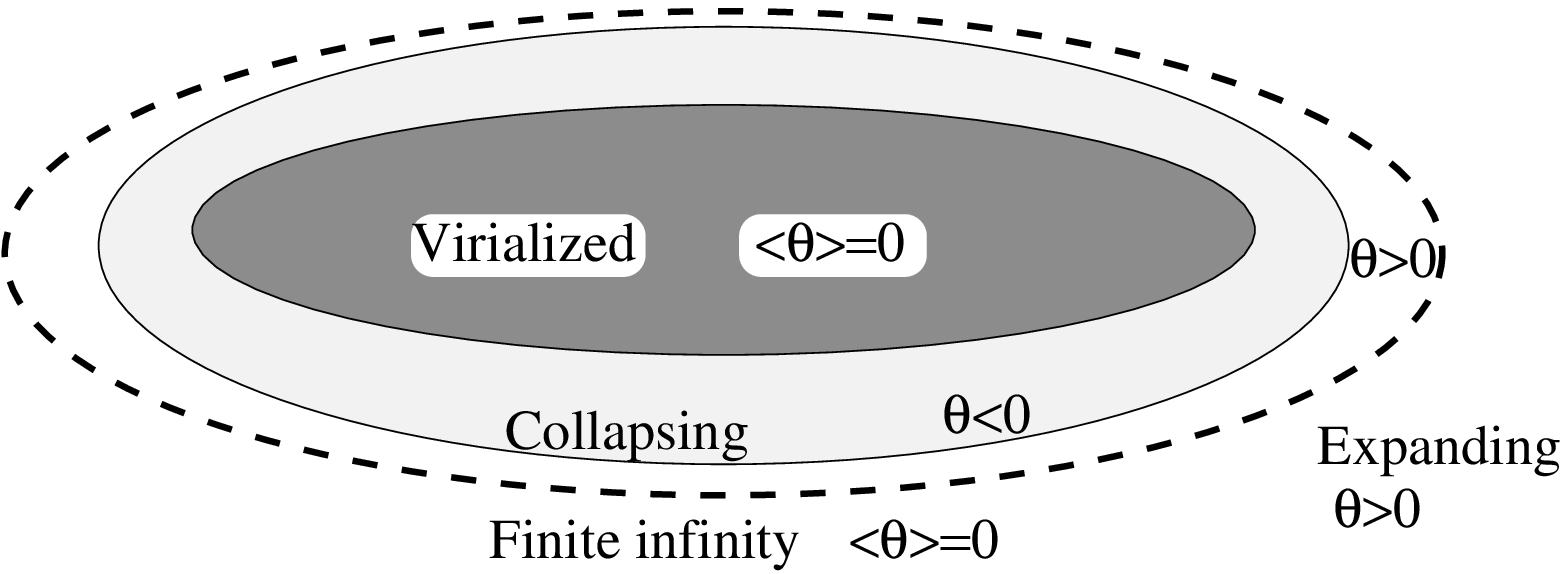,width=3.5in}}
\caption{Finite infinity, $\finfty$.}
\label{fifig}
\end{figure}
In view of the existence of backreaction, what is required is a notion of
true critical density without assuming evolution via the Friedmann equation.
This should be a dynamically evolving demarcation scale between systems which
will become bound and those that are unbound, given available data within the
past light cone at any epoch. As outlined in Ref.\ [\refcite{opus}],
and depicted in Fig.\ \ref{fifig}, finite infinity is defined in terms of
a scale over which the average expansion, including
virialized, collapsing and marginally expanding regions is zero,
$\ave\Th\!\Z{\Fi}=0$, while being positive outside.
Then $\rhcr\equiv\ave\rh\!\Z{\Fi}$. Finite infinity
is a non--static boundary analogous to the spheres cut out in the Swiss
cheese model\cite{gruyere}, but it involves average geometry rather
than matching exact solutions, and no assumptions about homogeneity
are made outside finite infinity. Finite infinity represents a physical
scale expected to lie outside virialized galaxy clusters, but within the
filamentary walls surrounding voids. Since it is a scale related to
the true critical density, space at finite infinity boundaries can be
described by the spatially flat metric
\beq\ds^2\Z{\Fi}=-\dd\tw^2+\aw^2(\tw)\left[\dd\etw^2+
\etw^2\dOM\right]\,,
\label{figeom}\eeq
where $\dOM\equiv\dd\th^2+\sin^2\th\,\dd\phi^2$.
While I use the term ``bubble wall'' it should be recognized that an
ideal bubble wall would consist of finite infinity regions touching
exactly at their boundaries. Due to the existence of minivoids one
should take care in identifying the scale of local walls empirically.

\section{Average homogeneity in a lumpy universe}

An important issue that has to be faced is the
split of space and time. Density is not a covariant quantity, but
will depend on the foliation of hypersurfaces
chosen to average on, and the specification of the observers within
the hypersurfaces. The hypersurfaces chosen by Buchert are the standard
ones based on surfaces of constant, $t$, where this is the affine
parameter on geodesics of ideal comoving observers. This choice, which
is analogous to the choice of synchronous gauge in FLRW models, is always
possible for an energy--momentum tensor with dust in the absence of
vorticity.

While it is entirely reasonable that Buchert's choice of gauge is
consistent at the volume--average position, provided one averages
over suitably large spatial volumes, there is a problem of physical
interpretation since observers in bound systems are located in
places where the physical assumptions in Buchert's gauge choice no
longer apply. As soon as regions start to collapse, geodesics will cross,
vorticity comes into play and comoving coordinates cannot be chosen
globally. Actual observers are in locations where the geometry is
better modelled by the vacuum Schwarzschild and Kerr solutions, rather
than by a rotationless dust fluid.

Since a nearly isotropic Hubble flow is observed, it is clear that
at some level a uniform Hubble flow should be obtained, despite the
large--scale inhomogeneity. I will make a choice of surfaces of
homogeneity which {\em implicitly solves the Sandage--de Vaucouleurs
paradox} by the assumption that on suitably small scales of averaging
{\em the bare Hubble flow is uniform below the scale of apparent
homogeneity},
\beq{1\over\ell_r(t)}\Deriv\dd t{\ell_r(t)}
=\frac13\ave\th\Z{\DD\X1}=\frac13\ave\th\Z{\DD\X2}=\cdots=\bH(t),
\label{homo}\eeq
$t$ being the local proper time in each averaging region $\DD\Z I$, and
$\ell_r\equiv\Vav^{1/3}=\bigl[\int_{\DD\X I}\dd^3x\sqrt{\gd}\bigr]
^{1/3}$ a relevant proper distance. Here no $\DD\Z I$ can be contained
within a finite infinity region. The important point is that
although it will appear to any one observer that voids expand faster
than bubble walls, the clocks of observers within voids will also tick
faster than those of observers at finite infinity on account of the
relative gravitational energy gradient.
Thus, provided the gravitational energy and spatial curvature gradients
are correlated, apparent variance in the Hubble flow for any one observer is
nonetheless perfectly consistent with uniformity of the bare, or
``quasilocally measurable'' Hubble flow.

The uniform expansion gauge, which is equivalent to one of the standard
gauges in cosmological perturbation theory\cite{Bard},
is given as an ansatz for defining the surfaces of homogeneity. However,
it should be noted that this ansatz is {\em less restrictive} than the
standard ansatz implicit in the FLRW models, where expansion is uniform, and
all ideal isotropic observers also have synchronous clocks and measure
the same spatial curvature on the relevant surfaces of homogeneity.
Here uniformity in the bare Hubble expansion is retained, but spatial
curvature and gravitational energy of isotropic observers vary in
a correlated fashion.

It should be emphasized that the Copernican principle is retained here.
If we compare ideal isotropic observers, some in galaxies and some in voids,
they will also measure an isotropic CMB. However, they will potentially
measure different mean CMB temperatures, and different angular scales
in the CMB anisotropies. These two observations respectively relate to
relative gravitational energy and relative spatial curvature. While
the clocks of all ideal observers are synchronized initially, when the
FLRW approximation was a good one, ultimately they will diverge; by
38\% on average at the present epoch. Since we exchange photons with
other bound systems, which are within finite infinity regions, such a
large clock--rate variance is not directly detectable in experiments
conceived to date.

The uniform bare Hubble constant gauge is more than simply an ansatz
which implicitly solves the Sandage--de Vaucoulers paradox while dealing with
the question of cosmological gravitational energy. As is argued in a separate
paper\cite{equiv}, it may be realized as a consequence of the strong
equivalence principle applied to expanding space. Among the class of all
possible motions of a timelike geodesic congruence there is a class of
conformal motions which
{\em do not isolate any direction in space as preferred}, in contrast to
individual boosts. The statement then is that for such motions we cannot
distinguish the circumstance in which the particles are {\em moving} from a
common origin in a static Minkowski space, or alternatively are {\em at rest}
in the underlying expanding universe. The two situations are equivalent;
giving a {\em Cosmological Principle of Equivalence}\cite{equiv}. In terms of
the conceptual basis of relativistic cosmology, this amounts to
a refinement of the Principle of Inertia. Not only can we not distinguish
which one of Galileo's ships\footnote{Actually Galileo just talked about a
single ship, in a context more like Einstein's closed elevator, but the
principle is the same.}, or of Einstein's trains is the one that is moving;
for conformal motions we can also not distinguish whether
it is the collection of ships or trains that are moving, or the `sea' or
`railyard' expanding in between.

\section{The fractal bubble model}
In Ref.\ [\refcite{opus}] I have written down a Buchert average of
the Einstein equations based on two scales, finite infinity regions
within which the geometry is given by (\ref{figeom}), and voids within
which the geometry is negatively curved with local scale factor, $\av$.
The geometry near the centres of voids is given by
\beq\ds^2\Z{\DD\X C}=-\dd\tv^2+\av^2(\tv)\left[\dd\etv^2+\sinh^2(\etv)
\dOM\right]\,.
\label{vogeom}\eeq
The Buchert equations are not written in terms of the local
geometry (\ref{figeom}) or (\ref{vogeom}) at either finite infinity
or the void centres, but in terms of an intermediate volume--average location
with a volume--average time parameter, $t$, and an average scale factor,
$\ab$, defined by
\beq\ab^3=\fvi\av^3+\fwi\aw^3,\eeq
$\fvi$ and $\fwi=1-\fvi$ being the respective initial void and wall
volume fractions at last scattering. Initially $\fvi\ll1$, and $\fwi\simeq1$,
since the universe is homogeneous and isotropic at last--scattering,
evolving like an Einstein--de Sitter one. The Buchert average is
constructed over the entire present particle horizon volume.

Although the actual surfaces of homogeneity will not coincide with
Buchert's ones on small scales,
we can still make use of Buchert's scheme, provided that we take care
in defining the relationship between the volume--average quantities of
Buchert's scheme, and those in a finite infinity region. Ultimately,
in order to deal with actual gradients in spatial curvature between
finite infinity and void centres, it may be necessary to use a scheme
such as that of Zalaletdinov \cite{Zal}, or to look at the problem
in terms of Ricci flow \cite{BC1}. This may be necessary, for example,
in determining the most appropriate gauge for the background to structure
formation simulations. The approach adopted in Refs.\
[\refcite{opus,sol,LNW}] is that the Buchert equations are solved in
volume average time, but the uniform expansion ansatz is used to
calibrate observable quantities relative to observers at finite infinity.
This involves a dressing of cosmological parameters over and above the volume
factors considered by Buchert and Carfora\cite{BC1,BC2}.

The independent Buchert equations, including the integrability
condition that ensures their closure, may be written
\bea
&&\OMM+\OMk+\OMQ=1,\label{Beq1}\\
&&\ab^{-6}\pt_t\left(\OMQ\bH^2\ab^6\right)+\ab^{-2}\pt_t\left(\OMk\bH^2\ab^2
\right)=0\,,\label{Beq2}
\eea
where the volume--average matter, curvature and kinematic backreaction
parameters are respectively
\beq \OMM={8\pi G\rhb\Z{M0}\ab\Z0^3\over3\bH^2\ab^3}\,;\
\OMk={-\kv\fvi^{2/3}\fv^{1/3}\over\ab^2\bH^2}\,;\
\OMQ={-\dot\fv^2\over9\fv(1-\fv)\bH^2}\,.\label{eOMQ}\
\eeq
The average curvature is due to the voids only, which are assumed
to have $\kv<0$, an overdot denotes a derivative w.r.t.\ volume--average
time, $t$, and $\bH\equiv\dot\ab/\ab$ is the volume--average or bare
Hubble parameter.

Although the quantity, $\bH$, has no particular significance for a general
Buchert average, for our particular assumptions it represents the
underlying uniform quasilocally measured Hubble flow. The quantities
$\OMM$, $\OMk$ and $\OMQ$ are then interpreted as the bare cosmological
parameters, relevant to a comoving isotropic observer at an average position
in freely expanding space, which will be within a void, but not at its
centre. These quantities are the closest analogues of the standard FLRW
density parameters.

It must be recalled that the scale factor $\ab$ does not correspond to
an exact geometry substituted into the Einstein equations,
and then solved. Rather we integrate the Buchert equations and then
reconstruct the average spherically symmetric geometry
\beq
\ds^2=-\dd\ts^2+\ab^2(\ts)\,\dd\etb^2+A(\etb,\ts)\,\dOM,
\label{avgeom}\eeq
the area function $A$ being defined by a horizon-volume average \cite{opus}.
The fact that (\ref{avgeom}) is spherically symmetric reflects the fact that
the geometry is reconstructed by taking an average on our radial null
geodesics. It is therefore {\em not} an LTB model, since exact spherical
symmetry has not been assumed in the Einstein equations.

Since (\ref{avgeom}) does not correspond to the local geometry of a
finite infinity observer, if we assume that our rods and clocks differ
little in calibration from those at finite infinity, we still have to be
careful in relating our local geometry (\ref{figeom}) to (\ref{avgeom}).
We do this by conformal matching of the radial null geodesics of
(\ref{figeom}) and (\ref{avgeom}), which requires that
$\dd\etw= \fwi^{1/3}\dd\etb/[\gb\fvf^{1/3}]$.
It then turns out that the finite infinity geometry may be rewritten
\beq \ds^2\Z{\Fi}=
-\dd\tw^2+{\ab^2\over\gb^2}\left[\dd\etb^2+\rw^2(\etb,\tw)\,\dOM\right]
\label{wgeom}\eeq
where
$\rw\equiv\gb\fvf^{1/3}\fwi^{-1/3}\etw(\etb,\tw)$. The geometry (\ref{wgeom}),
which now again has a more general spherically symmetric form, is
effectively the closest match to the FLRW geometry we usually attempt
to fit to the whole universe with the assumption that spatial curvature
everywhere matches our own, and clocks everywhere are synchronized to our
own wall time, $\tw$.

In place of the bare cosmological parameters (\ref{eOMQ}),
one can define conventional dressed parameters with respect to the geometry
(\ref{wgeom}), relevant to ``wall observers'' such as ourselves.
In particular, the dressed density parameter, $\OmM$,
defined according to $\OmM=\gb^3\OMM$ is effectively the conventional
parameter, whose numerical value will be
similar to that which we infer in FLRW models. Since (\ref{wgeom}) is not
an FLRW metric it does not make particular sense to supplement $\OmM$
by additional dressed parameters, as they will not sum to unity.

What is more significant cosmologically is that the dressed geometry
(\ref{wgeom}) will yield a dressed Hubble parameter,
\beq
\Hh\equiv{1\over a}\Deriv{\dd}{\tw}{a\hphantom{{}\ns w}}
=\gb\bH-\Dts\gb=\gb\bH-\gb^{-1}\Dtc\gb\,.
\label{42}
\eeq
where $a\equiv\ab/\gb$, which differs from the bare Hubble parameter,
$\bH$. Similarly, a wall observer will determine a dressed deceleration
parameter $q=-\ddot a/(\Hh^2 a)$, which differs from the bare
deceleration parameter $\bq=-\ddot\ab/(\bH^2\ab)$.

The general solution of the two scale Buchert equations was found in
Ref.\ [\refcite{sol}]. The general solution is specified by four independent
parameters, $\Hb$, $\OMMn$, $\fvn$ and $\epi$, which are respectively the
bare Hubble constant, the bare matter density, the present epoch void volume
fraction, and a small parameter $\epi\ll1$ related to the initial
void fraction. However, two of the four parameters
are greatly restricted by taking priors \cite{LNW} at
the surface of last scattering consistent with the CMB. Furthermore,
there is a tracker solution to which all solutions with these priors
approach to within 1\% by a redshift $z\goesas37$. This effectively
reduces the number of free parameters to two; we take these to
be $\Hb$ and $\fvn$. The tracker solution is given by
\bea
&&\ab={\ab\Z0\bigl(3\Hb t\bigr)^{2/3}\over2+\fvn}\left[3\fvn\Hb t+
(1-\fvn)(2+\fvn)\right]^{1/3}\\
&&\fv={3\fvn\Hb t\over3\fvn\Hb t+(1-\fvn)(2+\fvn)}\,,
\eea
in terms of the volume--average cosmic time, $t$. Since the
lapse function and bare Hubble parameter are related to the void
fraction by
$\gb=1+\half\fv=\frn32\bH t$,
the relation $\dd\tw=\gb^{-1}\dd t$ may be readily integrated to
obtain the wall time, $\tw$, relevant to observers in galaxies,
\beq
\tw=\frn23\ts+{2(1-\fvn)(2+\fvn)\over27\fvn\bH\ns0}\ln\left(1+{9\fvn\bH\ns0 t
\over2(1-\fvn)(2+\fvn)}\right).
\eeq

As is observed in Ref.\ [\refcite{sol}], expressions for many relevant
observable quantities, including effective dressed luminosity and angular
diameter distances, $\dL$ and $\dA$, are readily obtained. It is particularly
interesting to compare the bare volume--average tracker solution
deceleration parameter,
\beq\bq={2\fvf^2\over(2+\fv)^2},\eeq
with the corresponding dressed deceleration parameter
\beq
q={-\fvf(8\fv^3+39\fv^2-12\fv-8)\over\left(4+\fv+4\fv^2\right)^2}\,,
\label{qtrack}\eeq
At early times, when $\fv\ll1$, both deceleration parameters begin at
the Einstein--de Sitter value, $\bq\simeq q\simeq\half$. The bare
deceleration parameter always remains positive, meaning that a volume--average
observer in freely expanding space detects no cosmic acceleration, in
accord with the intuition of the critics\cite{IW}. However,
the dressed deceleration parameter changes sign at epoch when
$\fv\simeq0.5867$, at a zero of the cubic in (\ref{qtrack}), so that a wall
observer {\em does} detect apparent acceleration.

The fractal bubble (FB) model provides a solution to the
observational coincidence that the onset of cosmic acceleration appears to
roughly coincide with the growth of large structures.
Cosmic acceleration is an apparent effect that arises from gravitational
energy gradients related to spatial curvature gradients. It
begins when the void fraction
reaches 59\% at a redshift $z\simeq0.9$. This statement itself should be
able to be tested in future. Traditionally observational cosmologists
define voids in terms of a density contrast threshold, when quoting void
volume fractions\cite{HV}. For example, one might argue as to whether one
should take $\de\rh/\rh<-0.9$ or some other bound. In the present model, the
void volume fraction is related technically to those regions within the
past light cone which are not within finite infinity.
With some further refinement, it should be possible to relate this to
an observational definition of void fractions. Then as
surveys improve, the claims about the properties of $\fv(z)$ can be
compared with observation.

This illustrates the power of the FB model as compared to the spatially
flat \LCDM\ model. Both models depend on the Hubble parameter and one
other free parameter. However, $\Omega_\Lambda$ is not directly observable,
whereas the void--volume fraction, $\fv$, is empirically observable in
principle.

\section{Observational tests}

In Ref.\ [\refcite{LNW}] we have examined the fit of the FB model to
the Riess07 gold data set\cite{Riess06} of SneIa. We find that with
182 data points and two degrees of freedom the best--fit
$\chi^2=162.7$, i.e., a $\chi^2$ of approximately $0.9$ per degree
of freedom, which is a good fit.
While a marginally lower $\chi^2=158.7$
may be obtained for the best--fit flat \LCDM\ model (with $\Hm=62.6\kmsMpc$,
$\OmMn=0.34$, $\Omega\Z{\Lambda0}=0.66$), the difference is not significant.
In fact, a Bayesian model comparison of the FB model against a flat
\LCDM\ model with priors $55\le\Hm\le75\kmsMpc$, $0.01\le\OmMn\le0.5$.
gives a Bayes factor of $\ln B=0.27$ in favour of the FB model.
Statistically, such a small margin is ``not worth more than a bare
mention''\cite{KR} or ``inconclusive''\cite{Trotta}; that is to say,
the two models are statistically indistinguishable.
\begin{figure}
\centerline{\psfig{file=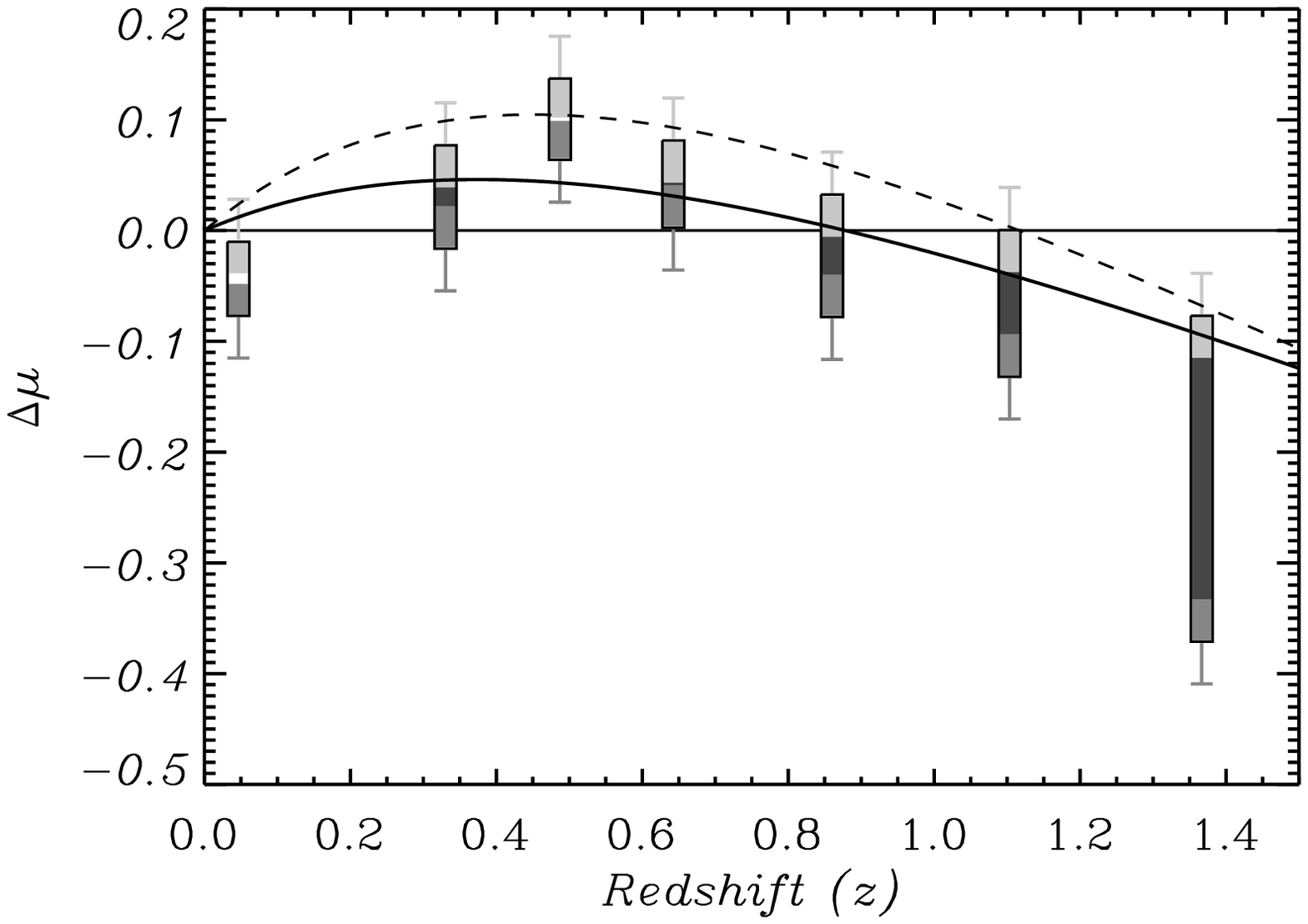,width=3.5in}}
\caption{The difference in the distance modulus, $\mu = 5 \log_{10} (\dL)
+ 25$, with $\dL$ in units Mpc, of the FB model with $\Hm=61.7\kmsMpc$,
$\OmMn=0.326$ from that of an empty coasting Milne universe, with the
same value of $\Hm$. The solid curve shows the FB model expectation, and
the dashed curve the expectation for a spatially flat \LCDM\ model with
the same values of $\Hm$ and $\Omega\Z{M0}$. Whiskers indicate how the
statistical uncertainties, shown as boxes, move when the background value of
$\Hm$ for the Milne universe which is subtracted is varied within the $2\si$
limits. For further details see Ref.\ [\refcite{LNW}].}
\label{dmu}
\end{figure}

The residual difference $\DE\mu=
\mu\Ns{FB}-\mu\ns{empty}$, in the standard distance modulus, $\mu=
5\log_{10}(\dL) + 25$, of the best--fit FB model
from that of a coasting Milne universe of the same Hubble constant,
$\Hm=61.7\kmsMpc$, is plotted in Fig.\ \ref{dmu} and compared
with binned data from the Riess07 gold data set. Apparent acceleration occurs
for positive residuals in the range, $z\lsim0.9$. (Note, however,
that the exact range of redshifts corresponding to apparent
acceleration also depends on the value of the Hubble constant of
the Milne universe used to compute the
residual.) The equivalent theoretical residual for the spatially flat \LCDM\
model with the same values of $\Hm$ and $\OmMn$ is also shown.
The difference in the gradients of the two curves should lead to observable
differences if observing programs which are trying to measure the
``equation of state, $P=w\rh$, $w(z)$, of dark energy'' can achieve sufficient
accuracy.
\begin{figure}
\leftline{\hskip-10pt\psfig{file=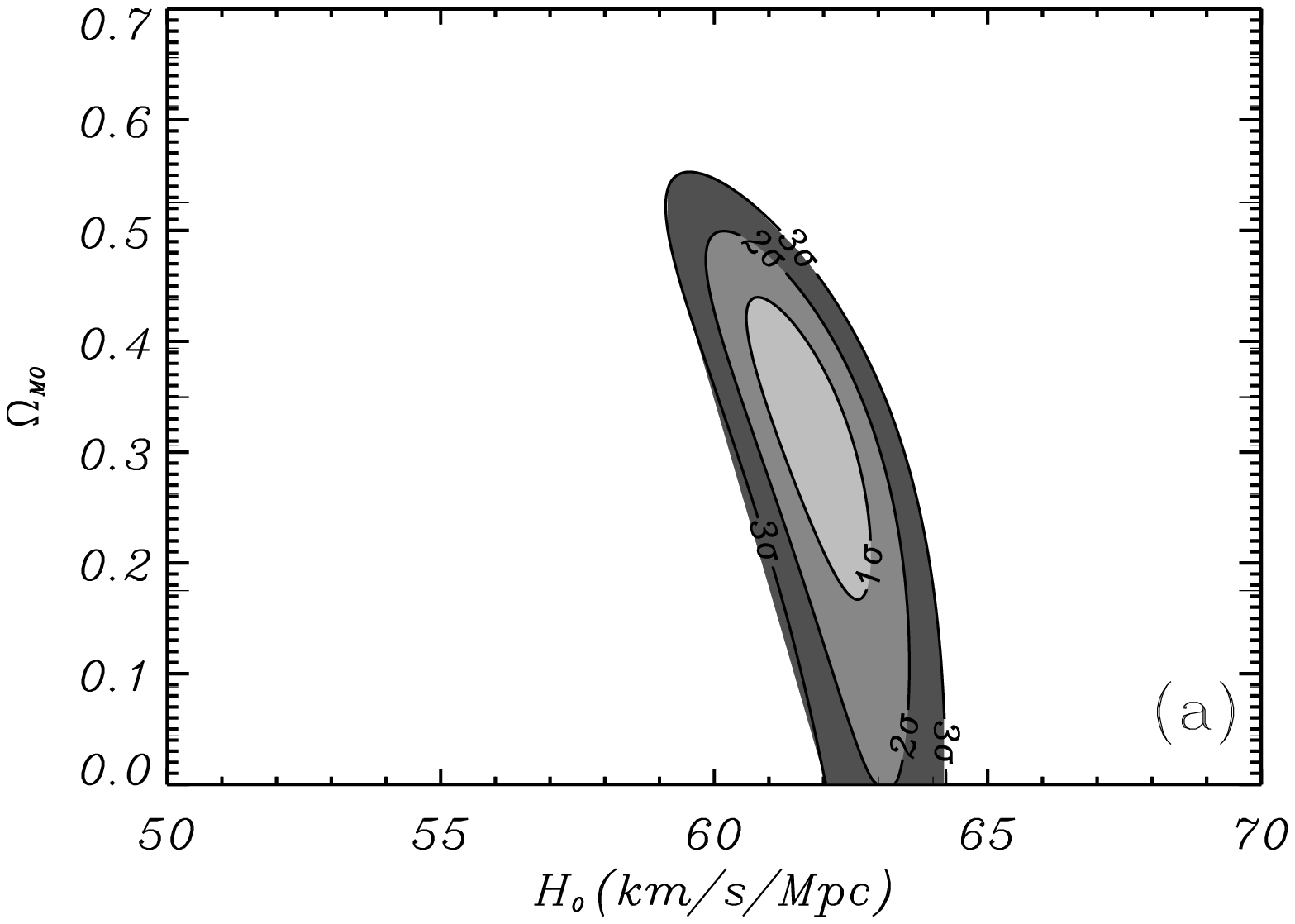,width=2.4in}\hskip-10pt
\psfig{file=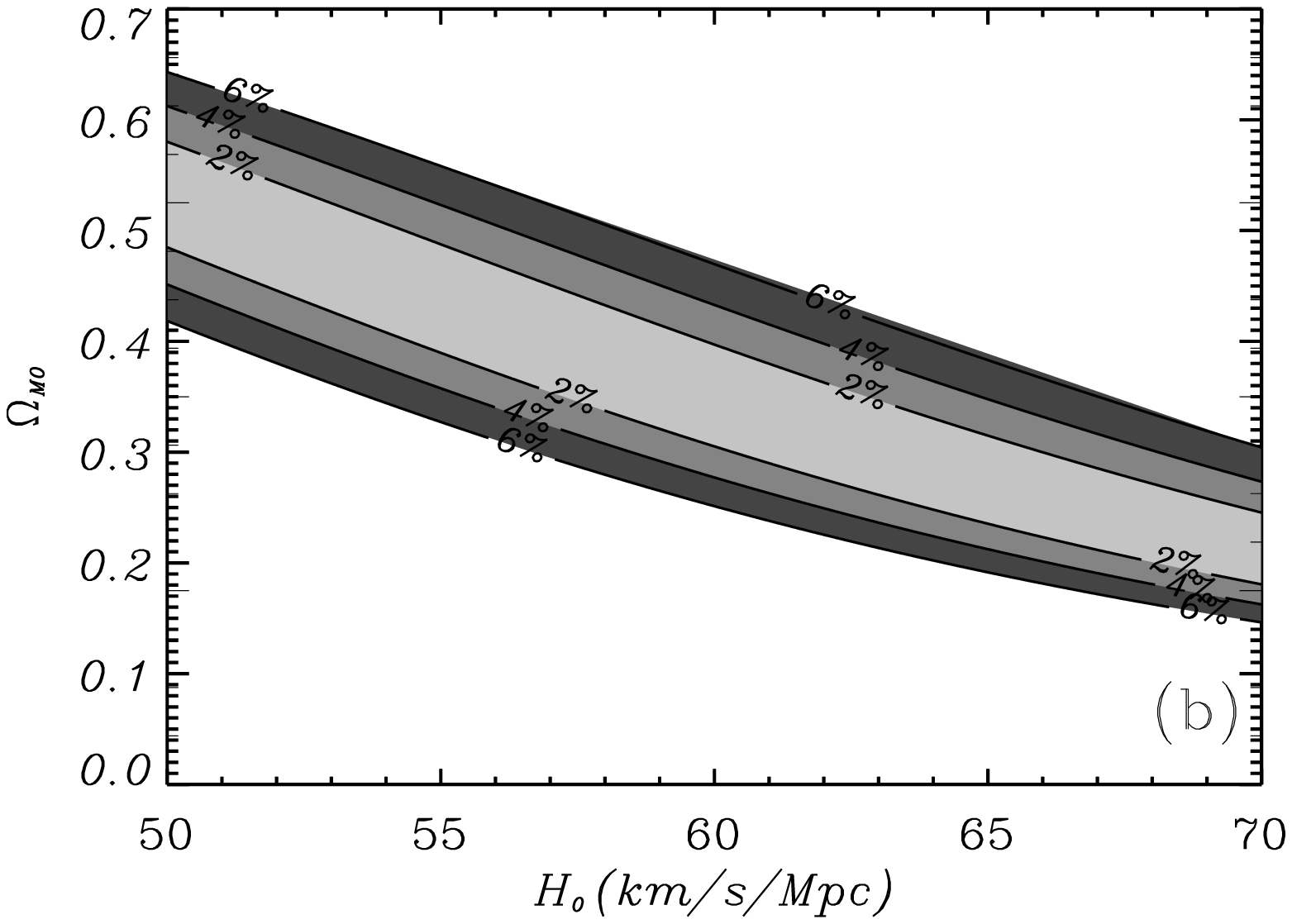,width=2.4in}}
\leftline{\hskip-10pt\psfig{file=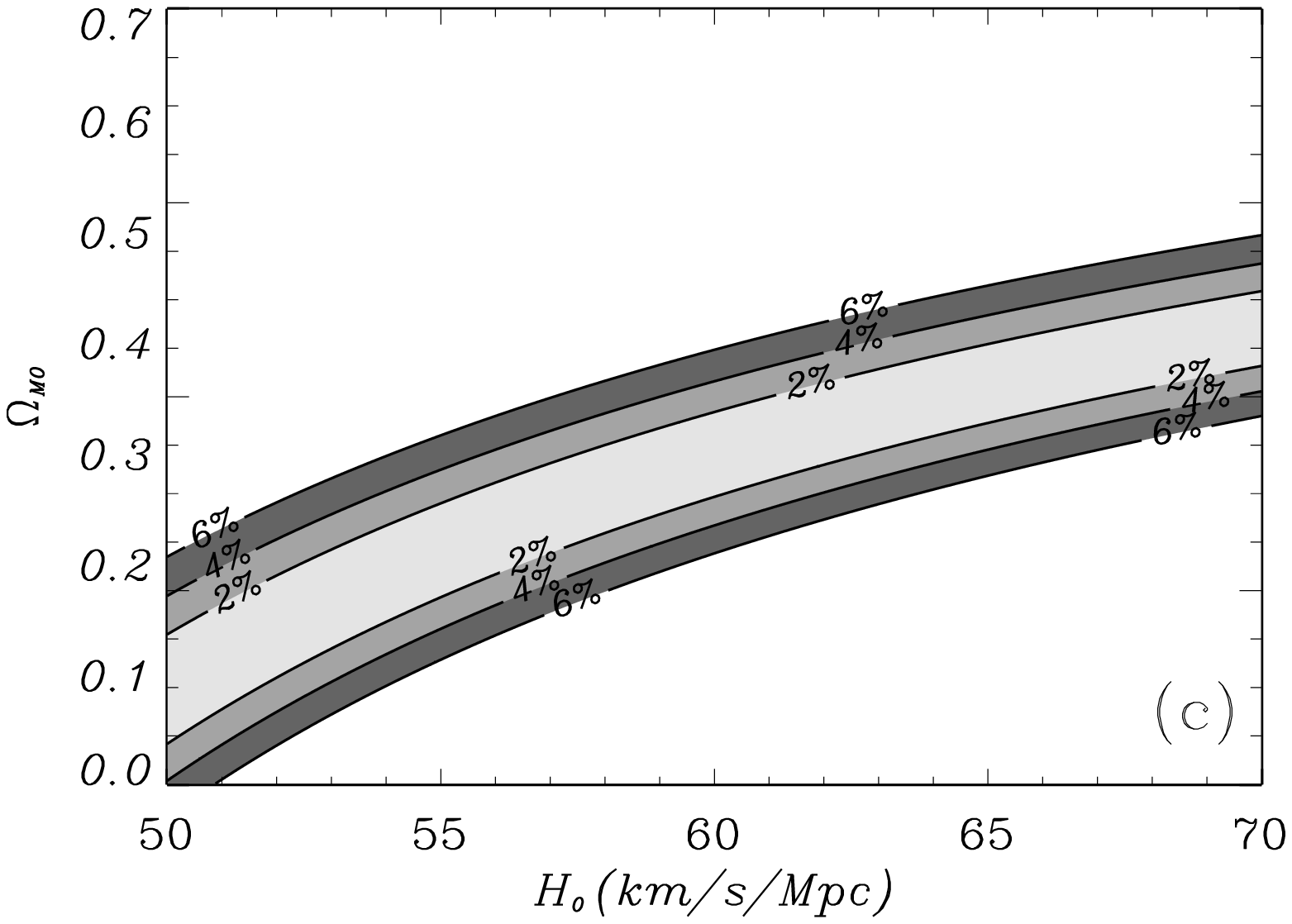,width=2.4in}\hskip-10pt
\psfig{file=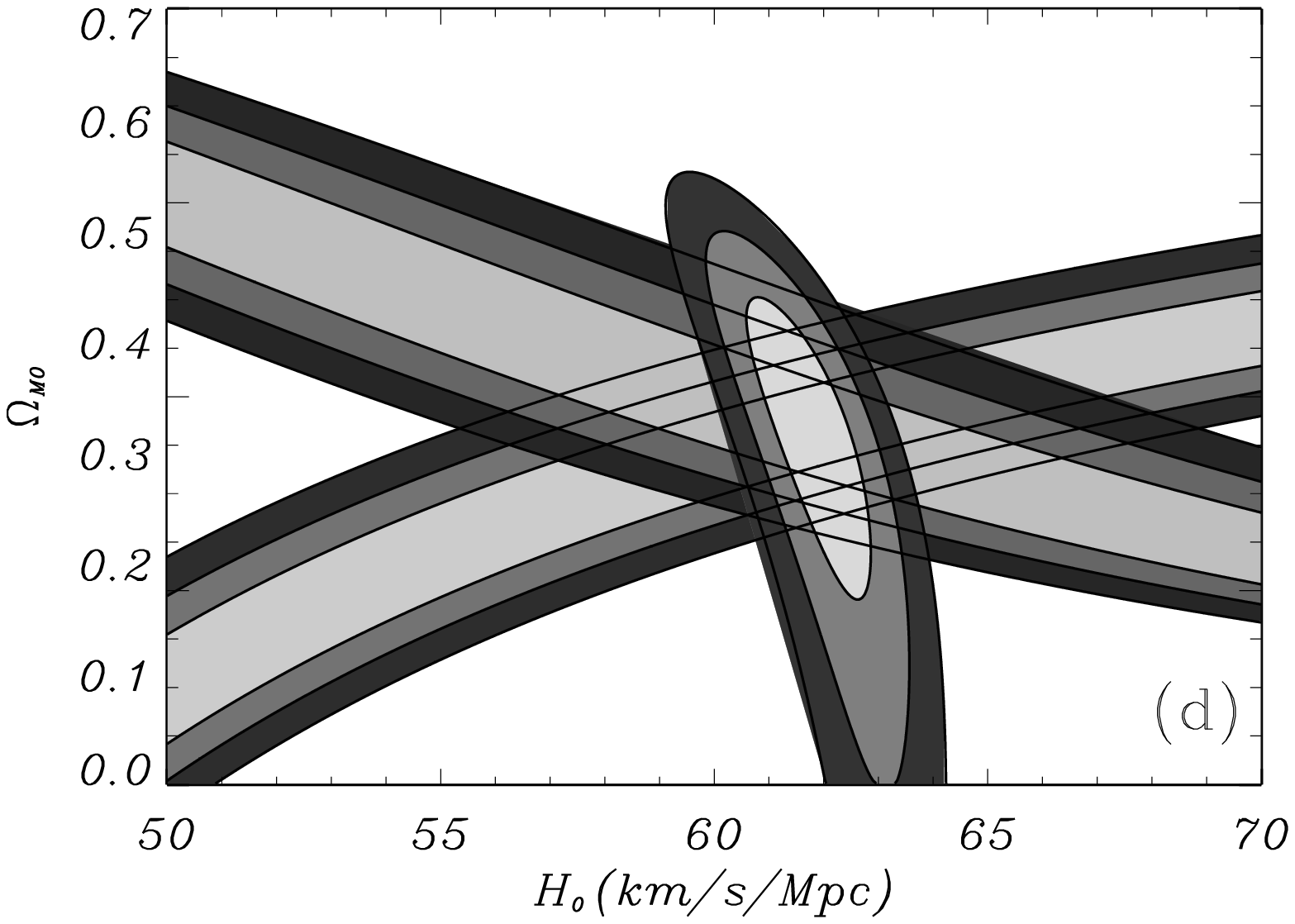,width=2.4in}}
\caption{Various cosmological tests in the parameter space of the dressed
Hubble constant, $\Hm$, and the dressed matter density parameter, $\OmMn$.
{\bf(a)} Statistical confidence limits on SneIa in the Riess07 gold data
set\cite{Riess06}; {\bf(b)} Parameters which match $\de=0.01$ angular scale
of sound horizon\cite{wmap1,wmap3} of \LCDM\ model to within 2\%, 4\%, 6\%;
{\bf(c)} Parameters which match effective comoving $104h^{-1}$ Mpc scale
of baryon acoustic oscillation\cite{bao} to within 2\%, 4\%, 6\%;
{\bf(d)} Overlay of panels (a),(b),(c).}
\label{tests}
\end{figure}

In Fig.\ \ref{tests} we address three major independent cosmological
tests. In addition to statistical confidence limits in the $(\Hm,\OmMn)$
parameter space we also consider the angular scale of the sound horizon
in the CMB anisotropy spectrum\cite{wmap1,wmap3}, and the comoving scale of
the baryon acoustic oscillation (BAO) seen in galaxy clustering\cite{bao}
using the relevant dressed geometry (\ref{wgeom}).

Ideally we should recompute the spectrum of Doppler peaks for the FB
model. However, this is a mammoth task, as the standard
numerical codes have been written solely for FLRW models, and every step
has to be carefully reconsidered. For this reason we first
ask whether parameters exist for
which the effective angular diameter scale of the sound horizon
matches the angular scale of the sound horizon, $\de=0.01$ rad, of the
\LCDM\ model, as determined by WMAP\cite{wmap1}. Since there is no
change to the physics of recombination, but just an overall change
to the {\em calibration} of cosmological parameters, this is
entirely reasonable. In Fig.~\ref{tests} we plot parameter ranges which
match the $\de=0.01$ rad sound horizon scale to within 2\%, 4\% and 6\%,
using the calculation of the sound horizon given in Ref.\ [\refcite{opus}],
Sec.\ 7.2. The 2\% contour
would roughly correspond to the 2$\si$ limit if the WMAP uncertainties for
the \LCDM\ model are maintained. As this can only be confirmed by detailed
computation of the Doppler peaks, the additional levels have been chosen
cautiously.

In the case of the BAO scale, as we do not yet have the
resources to analyse the galaxy clustering data directly, we also begin with
a simple but effective check. Since the dressed geometry
(\ref{wgeom}) does provide an effective almost--FLRW metric adapted to our
clocks and rods in spatially flat regions,
the effective comoving scale in this geometry should match the
corresponding observed BAO scale of $104h^{-1}$Mpc.
In Fig.\ \ref{tests} we therefore also plot parameter values which
match this scale to within 2\%, 4\% or 6\%.

It should be noted that even if SneIa
are disregarded, the parameters which fit the two independent
tests relating to the sound horizon and the BAO scale agree with each
other, to the accuracy shown, for values of the Hubble constant which include
the value of Sandage \etal\cite{Sandage}; but {\em not} for the values
of $\Hm$ greater than $70\kmsMpc$ which best--fit the WMAP
data\cite{wmap1,wmap3} with the FLRW model. The value $\Hm=62.3\pm
1.3\w{(stat)}\pm 5.0\w{(syst)}\kmsMpc$
determined by the Hubble Key Team of Sandage \etal\cite{Sandage}
has been controversial, given the 14\% difference from values which
best--fit the WMAP data with the \LCDM\ model\cite{wmap1,wmap3}. However,
the WMAP analysis only constitutes a direct measurement of CMB temperature
anisotropies; the determination of cosmological parameters involves model
assumptions. In the FB model, the model assumptions are different. In
particular, on account of the differences in gravitational energy and
local spatial curvature between observers in bound systems, and those
at the volume average in freely expanding space, the calibration of
quantities related to the CMB has to be carefully reconsidered. Observers
at the volume average detect a cooler mean CMB temperature\cite{opus}
$\bT\Z0=\gc^{-1}T\Z0=1.98$ K than the $T\Z0=2.73$ K
we measure. Accounting for these differences leads to concordance
for parameter values which include
the Hubble constant of Sandage \etal\cite{Sandage}.
\begin{table}
\tbl{Best--fit cosmological parameters derived from the independent
parameters, $\Hm$, $\fvn$.}
{\begin{tabular}{@{}lll@{}}\toprule
dressed Hubble constant & $\Hm=61.7^{+1.2}_{-1.1}\kmsMpc$\\
present void volume fraction & $\fvn=0.76^{+0.12}_{-0.09}$\\
mean lapse function & $\gc=1.381^{+0.061}_{-0.046}$\\
bare density parameter & $\OMMn=0.125^{+0.060}_{-0.069}$\\
conventional dressed density parameter & $\OmMn=0.33^{+0.11}_{-0.16}$\\
mass ratio of & \\
\quad non--baryonic dark matter to baryonic matter &
 $(\OMMn-\OMBn)/\OMBn=3.1^{+2.5}_{-2.4}$\\
bare Hubble constant&$\Hb=48.2^{+2.0}_{-2.4}\kmsMpc$\\
effective dressed deceleration parameter&$q\Z0=-0.0428^{+0.0120}_{-0.0002}$\\
age of universe measured in a galaxy&$\tau\ns{w0}=14.7^{+0.7}_{-0.5}$ Gyr.\\
\botrule
\end{tabular}
}\begin{tabnote}
$^{\text a}$ 1 $\sigma$ statistical uncertainties from SneIa alone
are shown.\\
\end{tabnote}
\label{tab1}
\end{table}

Best--fit cosmological parameters
are shown in Table \ref{tab1}. As statistical bounds
are not available for the sound horizon and BAO tests, we show 1$\si$
uncertainties from SneIa only. However, given that there are many independent
estimates of the dressed matter density, $\OmMn$, we expect that the
uncertainties quoted can be significantly reduced if such priors are
imposed.

\subsection{Resolving the lithium abundance anomaly}
The angular scale of the sound horizon and the BAO tests have
been applied assuming a {\em volume--average} baryon--to--photon ratio
in the range $\etBg=4.6$--$5.6\times10^{-10}$ adopted by
Tytler \etal\cite{bbn2} prior to the release of WMAP1. With this range it
is possible to achieve concordance with lithium abundances, while
also better fitting helium abundances. This may resolve the primordial
lithium abundance anomaly\cite{bbn1,lithium}.

With the 2003 WMAP1 release\cite{wmap1}, the
baryon--to--photon ratio was increased to the very upper range of values
that had previously been considered, due to its
effect on the ratio of the heights of the first two Doppler
peaks. This ratio of peak heights is sensitive to the mass
ratio of baryons to non--baryonic dark matter -- rather
than directly to the baryon--to--photon ratio -- as it depends physically on
baryon drag in the primordial plasma. The fit to the Doppler peaks
required more baryons than the range of Tytler \etal\cite{bbn2} admitted, when
calibrated with the FLRW model. In the FB calibration, on account of the
difference between the bare and dressed density parameters, a bare value
of $\OMBn\simeq0.03$ nonetheless corresponds to a conventional dressed
value $\Omega\Z{B0}\simeq0.08$, and an overall mass ratio of
baryonic matter to non--baryonic dark matter typically of
about 1:3, which is larger than for \LCDM.
This would certainly indicate sufficient baryon drag to
accommodate the ratio of the first two peak heights.

\subsection{Spatial curvature and the ellipticity anomaly}
Since the release of the Boomerang experiment\cite{boom} in 2000, it
has been generally assumed that the angular position of the first Doppler
peak is a measure of the spatial curvature of the universe. However, this
inference assumes that the spatial curvature is the same everywhere,
as is appropriate for the FLRW models.
In the present model there are spatial curvature gradients, and we must
revisit the calculation from first principles, as outlined in Ref.\
[\refcite{opus}]. Insofar as the angular scale of the sound horizon
is reproduced for the FB calibration, the overall angular scale cannot
be regarded as a measure of average spatial curvature.

If there is a sizable negative average spatial curvature at
the present epoch, then there must be ways of detecting it other than
via the measurement of the angular scale of the Doppler peaks.
Such effects do indeed arise when one considers more subtle measurements
associated with the average geodesic deviation of null geodesics.
Indeed one prediction is that there should be nontrivial ellipticity
in the CMB anisotropies on account of greater geodesic mixing.
This effect is in fact observed\cite{ellipticity}, and is an important
anomaly for the standard \LCDM\ paradigm, overlooked by the majority of the
community to date.

A detailed quantification of the degree of ellipticity expected, and
its comparison with the observed signal, will be an important
test of the FB model in future.

\subsection{The WMAP3 normalization}
It has been recently noted that the values of the normalization of the
primordial spectrum $\si\Z8\goesas0.76$ and matter content $\OmMn\goesas0.24$
implied by WMAP3 are barely compatible with the abundances of massive clusters
determined from X--ray measurements\cite{omegam}. In fact, SneIa also best
fit the \LCDM\ model with higher values of $\OmMn$ comparable to the the
best--fit value for the FB model, namely $\OmMn\goesas0.33$. In the FB model
the best--fit void fraction, $\fvn\goesas0.76$, appears to be the quantity
that is mimicked by dark energy fraction, $\Omega\Z{\Lambda0}$, as far as the
WMAP normalization to the \LCDM\ model
is concerned. The flat \LCDM\ model constrains $\OmMn=1-\Omega\Z{\Lambda0}$.
For the FB model, the corresponding constraint, $\OmMn\simeq\frn12(1-\fvn)
(2+\fvn)$, gives quite different larger values, consistent with many other
observational determinations of the conventional matter density parameter.

\subsection{The expansion age}
Structure formation scenarios in standard \LCDM\ model have some
difficulty in explaining the observed apparent very early formation of
galaxies \cite{red}. Of course, structure formation contains many
model--dependent assumptions, so direct measurements of the ages of
metal poor stars in old globular clusters are ultimately a better
indicator. The observational bounds are not very tight, because
of the large uncertainties involved in particular nuclear processes.
Individual ages are generally consistent with the accepted 13.7Gyr
\LCDM\ age of the universe, but are sometimes in tension with it\cite{oldies},
given our uncertainties in knowing how early on the first stars could form.

It is interesting to note that the FB model not only adds a billion
years to the age of the universe as measured in a galaxy, but that the
percentage difference is larger at earlier times. For concordance
\LCDM\ $\tau=0.85$ Gyr at $z=6.4$ when distant quasars are seen, and
$\tau=0.365$ Gyr at reionization at $z=11$. By contrast for the best--fit
FB model $\tau=1.14$ Gyr at $z=6.4$ and $\tau=0.563$ Gyr at $z=11$,
making the universe 34\% and 54\% older than concordance \LCDM\
at the respective epochs.

The fact that the expansion age of the universe determined in a galaxy is
greater than that of concordance \LCDM\ may alleviate but
not completely solve various aspects of the age problem. However, the
fact that the age of the universe is 18.6Gyr at the volume average is
a basic indicator that the whole issue of structure formation,
including calibrations, needs to be re-examined from first principles.

\subsection{The Hubble bubble}
Recent analysis of SneIa data by Jha, Riess and Kirshner \cite{JRK} confirms
an effect which has been known about for some time, and has been interpreted
as our living in a local void \cite{Tom1}, the ``Hubble bubble''. If one
excludes SneIa within the Hubble bubble at redshifts
$z\lsim0.025$, then the value of the Hubble constant obtained is lower by
$6.5\pm1.8$\% \cite{JRK}. It is for this reason that SneIa at $z\le0.023$ have
been excluded from the Riess07 gold data set\cite{Riess06}. The Hubble bubble
is very problematic for the standard dark energy paradigm, as it can
contribute a systematic error to the estimate of the equation of state
parameter, $w$, which is much larger than the statistical uncertainties
that precision cosmologists hope to attain\cite{JRK,essence}.

In the FB model the Hubble bubble is expected as a feature that results
from the implicit resolution of the Sandage--de Vaucouleurs paradox.
Since the bare Hubble parameter characterizes the uniform ``quasilocally
measured'' Hubble flow, eq.~(\ref{42}) also quantifies the apparent variance
in the Hubble flow below the scale of homogeneity. The present bare Hubble
constant, $\Hb$, is lower than the global average, $\Hm$. It represents the
value we as observers in galaxies would obtain for measurements
averaged solely within the plane of an ideal local bubble wall, on scales
dominated by finite infinity regions. For us in particular, measurements
of the Hubble constant towards the Virgo cluster would represent a scale
over which we might hope to detect the bare Hubble constant, $\bH\goesas
48\kmsMpc$, as this appears to be the direction that most closely represents
a local bubble wall. Ultralocal minivoids\cite{minivoids} complicate
any empirical measurement.
\begin{figure}
\centerline{\psfig{file=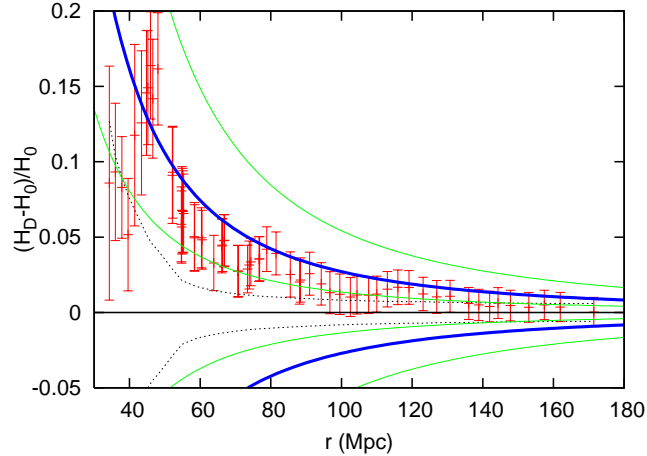,angle=270,width=3.5in}}
\caption{Scale dependence of the normalized difference of the Hubble rate,
averaged within a domain, $H\Z D$, and the global average, $\Hm$, with
data of Freedman \etal\cite{Freed}. Courtesy of Li and Schwarz\cite{LS2},
who provide further details.}
\label{bubble}
\end{figure}

`Local' measurements across single voids of the dominant $30h^{-1}\w{Mpc}$
diameter\cite{HV}, probe the scale over which photons on null geodesics
encounter the fewest finite infinity domains.
Such measurements should give a Hubble `constant' which exceeds the global
average $\Hm$ by an amount commensurate to $\Hm-\Hb$. As voids are dominant
by volume, an isotropic average will generally produce a Hubble `constant'
{\em greater than} $\Hm$ for local averages until we sample on large
enough scales that the volume average of walls and voids is the same as
the global one. Thus the average
will steadily decrease from its maximum at $\goesas30h^{-1}$ Mpc
until the scale of homogeneity ($\goesas100h^{-1}\w{Mpc}$) is reached: the
Hubble bubble feature. This expectation is dramatically confirmed
by the data points in Fig.~\ref{bubble}, reproduced from a recent paper of
Li and Schwarz\cite{LS2}. With $h=0.617$, the maximum Hubble
`constant', up to about 20\% higher than average, should be attained at a
scale of about 48Mpc, thereafter steadily decreasing until leveling out at the
scale of homogeneity, which must be reached before the BAO scale of about
168 Mpc.

Fig.\ \ref{bubble} also illustrates how the resolution of the Sandage-de
Vaucouleurs paradox gives unique predictions. The two thick lines in Fig.\
\ref{bubble} show Li and Schwarz's estimates of the bounds between which
the data should lie using Buchert averaging, but not accounting for the
clock effect.
Given that the data indicates a consistently higher Hubble constant below the
scale of homogeneity, the only explanation for the lack of great statistical
scatter between the two lines -- if all clocks are effectively synchronous
-- is that by a statistical fluke we happen to be in a large local void which
is expanding faster into the surrounding medium\cite{LTBmatch,Tom1}. However,
if the conventional FLRW assumptions apply beyond this Hubble bubble,
then by the standard structure formation scenarios such a circumstance
seems impossible.

In the FB model the Hubble bubble is not a statistical fluke, but
like cosmic acceleration is an apparent effect that arises from
clock rate variance. Distant observers in galaxies beyond our
own Hubble bubble will also detect a Hubble bubble centred on their
location. The theoretical derivation of the curve that should pass
though the data points in Fig.\ \ref{bubble} will provide an important test
of the FB model. To determine this curve, and its variance, requires
some knowledge of void statistics, and probably some Monte--Carlo
simulations of the manner in which the scale of homogeneity is filled
on average. Particular anisotropies which look like ``large''
voids\cite{bigvoid} might be expected from the variance in alignments
of dominant $30h^{-1}$ diameter voids below the scale of homogeneity.
A systematic error of 1--2\% in the CMB dipole subtraction\cite{Freeman}
may well be the consequence of a Rees--Sciama dipole resulting from
such foreground inhomogeneities\cite{RRS}.

While the decades long debate among astronomers about
the value of the Hubble constant has been mainly dominated by arguments about
systematics\cite{Sandage,Feast}, the question of the scale of averaging has
no doubt also played a role. Lower values of $\Hm$ will be expected if one
specifically selects directions and scales which approximate our own bubble
wall\cite{virgo}. However, distance determinations on scales
$\lsim50$ Mpc will generally give higher values of $\Hm$. Since many of the
first steps on the cosmic distance ladder are calibrated on such scales, many
related issues require careful reconsideration.

\subsection{Prospects for future cosmological tests}

The FB model provides a strong new competitor to the standard \LCDM\
cosmology, as Table \ref{concord} shows. Indeed, since it steps out
of a paradigm in which everything rests on a single
equation of state, it will provide new and unique predictions. Current
observational programs of course focus on the measurement of
$w(z)$. The difference in expectations of the FB model for those
programs will be highlighted in a forthcoming paper\cite{obs}.
\begin{table}
\tbl{Model comparison}
{\begin{tabular}{@{}lll@{}}\toprule
Observation&\LCDM&Fractal bubble model$^{\text a}$\\
\toprule
SneIa luminosity distances&Yes &Yes\\
BAO scale (clustering)&Yes &Yes\\
Sound horizon scale (CMB) &Yes &Yes\\
Doppler peak fine structure &Yes &Still to calculate\\
Integrated Sachs--Wolfe effect &Yes &Still to calculate\\
Primordial $^7$Li abundances &No &Yes\\
CMB ellipticity &No &[Yes]\\
CMB low multipole anomalies &No &[Foreground void(s):\\
&&$\hphantom{[}$Rees--Sciama dipole]\\
CMB Sunyaev--Zel'dovich signal\cite{SZ1,SZ2} &No &Still to calculate\\
Hubble bubble &No &Yes\\
Nucleochronology of globular clusters &Tension &Yes\\
X-ray cluster abundances &Marginal &Yes\\
Emptiness of voids &No &[Yes]\\
Sandage-de Vaucouleurs paradox &No &Yes\\
Coincidence problem &No &Yes\\
\botrule
\end{tabular}
}\begin{tabnote}
$^{\text a}$ Square brackets indicate cases where
the present indication is suggestive, but much detailed calculation
remains to be done.\\
\end{tabnote}
\label{concord}
\end{table}

The quantities that remain to be calculated in Table \ref{concord}
relate largely to the CMB. Detailed construction of the Doppler peaks,
to enable comparison to WMAP, and determination of parameters such as
$\si\Z8$, is a matter of urgency. It is a rather
non--trivial exercise in recalibration of standard quantities, in which
all steps need to be carefully reconsidered. Detailed quantitative
calculations of CMB ellipticity and the integrated Sachs--Wolfe effect
can only be performed in conjunction with such an analysis.

While the calculation of the integrated Sachs--Wolfe (ISW) effect will differ
in the FB model, one must be careful not to base intuition on that of
perturbations on FLRW backgrounds. In particular, the observed ISW signal is
believed to confirm dark energy since its consequence is a suppression
of the gravitational collapse of matter at relatively recent times\cite{isw}.
If one replaces the words ``dark energy'' by ``voids'' in the standard
qualitative explanation, then a probable description of the ISW effect
in the FB model emerges. The same correlation of the ISW signal with
clumped structures\cite{isw} is expected; what is important is the magnitude
of the effect. Mattsson\cite{Mat} has recently proposed an extension
of the Dyer--Roeder formalism, which may have some relevance for such
calculations.

In offering a new paradigm for cosmology, every observational test naturally
has to be revisited from first principles. Thus there are potential tests
which relate to both strong and weak gravitational lensing, for example,
which have not been listed in Table \ref{concord}. Qualitatively one
expects voids to be emptier in the FB model than in LCDM structure
formation simulations. However, the specification of the background
average surfaces of homogeneity, in terms of a uniform bare Hubble
expansion gauge, and a correct post--Newtonian approximation on such
a background, need to be examined before structure formation simulations
are attempted.

Ultimately, apparent variance in the Hubble flow below the scale of
homogeneity will give predictions which cannot be reproduced in the FLRW
scenario. Not only should we determine the average curve that passes
through the data points in Fig.\ \ref{bubble}, we should ultimately
collect thousands of SneIa measurements or other distance measurements
on scales up to 200Mpc, and test the correlation of the apparent Hubble
flow with actual structure: a sort of Hubble flow tomography of the
nearby universe.

\section{Conclusion}

A true ``concordance cosmology'' should agree with all reliable observations,
and not just a carefully selected subset. A glance at Table \ref{concord}
reveals that there are many anomalies in the standard \LCDM\ model.
Much attention has been focused on the CMB anisotropies, on account of
the spectacular success of the WMAP mission\cite{wmap1,wmap3}. However,
it must be recalled that many ingredients go into the analysis of the
CMB temperature fluctuations. In looking at these ingredients we must
ask {\em``what is the weakest link''}?

In my view, the weakest link is not primordial nucleosynthesis\cite{bbn1},
which is based on nuclear physics that we understand very well, but the
cosmological model. The weakest link comes from abandoning the
theoretical principles that we are careful to apply in other circumstances.
In particular, in general relativity one should model the universe with the
matter distribution one observes, rather than trying to impose onto the
universe a simple mathematical solution based on some simplifying
symmetry.

It is unfortunate that general relativists have been obsessed by
exact solutions of Einstein's equations, whether they involve likely or
unlikely approximations for the matter distribution. We should face up to
the fact that the solution for the actual matter distribution is
analytically intractable, and therefore the question of cosmological
averaging\cite{buch1,brpapers,fit1,fit2,Zal,CPZ,PS} is paramount.
Furthermore, once we do take an
average we must address the fundamental problem that the relationship
of rods and clocks at one point to those
at a distant point, a conceptual centrepiece of general relativity,
is highly non--trivial once gradients in spatial curvature and
gravitational energy are considered. In an expanding universe these involve
subtle dynamical aspects of general relativity, which cannot be localized
at a point on account of the equivalence principle.

Some colleagues when presented with Table \ref{concord}, and the
fact that the right hand column is based entirely on general relativity
with no new or exotic physics (beyond a need for non--baryonic dark
matter), suggest I should invoke Ockham's Razor at this point and declare
the cosmological constant dead. Other colleagues, who sometimes confess
to Newtonian intuition, are of the view that a clock--rate difference
of 38\% accumulated between bound systems and the volume average over the
lifetime of the universe is nonetheless so great that I must be mistaken
somewhere, in spite of Table \ref{concord}.

In my view caution should always be exercised, but this includes caution
with the conceptual basis of our theory and the operational interpretation
of measurements. To those who are uncomfortable with my proposal
about cosmological quasilocal gravitational energy let me ask the
following: Without reference to an asymptotically flat static reference
scale, which does not exist given the universe is expanding, and without
reference to a background which evolves by the Friedmann equation at some
level\cite{gruyere}, an assumption which is manifestly violated by the
observed inhomogeneities, {\em What keeps clocks synchronized in cosmic
evolution}? Please explain.

In the FB proposal the fact that our FLRW approximation has
served so well is understood as a consequence of the fact that the bare
Hubble flow is uniform, despite large--scale inhomogeneities. Uniformity
of the quasilocally measured Hubble flow does not imply uniformity of spatial
curvature, nor of gravitational energy, however. If we examine the Hubble flow
over scales on which the average gradients are not large, we will not see
large statistical scatter in the Hubble flow. But averaged over larger scales,
below the scale of homogeneity, we will see a variance in the apparent
Hubble flow, that agrees with observations of the statistical properties
of voids\cite{HV}, and which seems to accord with Fig.\ \ref{bubble}. The
fact that there is a statistically dominant void scale\cite{HV} may well be
associated with the evolution of the scale corresponding to the second
(rarefaction) Doppler peak in the CMB anisotropy spectrum, just as
evolution of the first (compression) peak appears to statistically define
the cutoff for the scale of homogeneity.

Any proposal which seeks to genuinely shift the paradigm against accepted
intuition, even a conservative proposal based on a principled
approach\cite{opus,equiv} to our best theory of gravitation, naturally faces
much initial scepticism. Nonetheless, I suspect that in future
cosmologists will wonder how we ever could have expected that two decades
of precision measurements would simply confirm the na\"{\i}ve cosmological
models constructed in the 1920s, long before we knew what the universe
actually looked like. I suspect that the coming decades will see us
explore and finally understand the arcane territory of general relativity
relating to quasilocal gravitational energy\cite{ge1,quasi,CLN,Ga}, informed
directly by new cosmological observations.\medskip

\noindent {\bf Acknowledgements} I am grateful to many people for discussions
and correspondence including, in particular, Thomas Buchert. I also thank
Anthony Fairall and Dominik Schwarz for granting me permission to reproduce
their Figures \ref{fig1} and \ref{bubble} respectively.


\end{document}